\documentclass[aps,prx,reprint,superscriptaddress]{revtex4-1}

\usepackage{graphicx}
\usepackage[export]{adjustbox}
\usepackage[normalem]{ulem}
\usepackage{natbib}
\usepackage{amsmath,array,amssymb}
\usepackage{dcolumn}
\usepackage{bm}
\usepackage{hyperref}
\usepackage{color}
\usepackage{amsmath, amsthm, amssymb, amsfonts}
\usepackage{todonotes}

\newcommand{\ket}[1]{\vert#1\rangle}
\DeclareMathOperator{\sech}{sech}
\newcommand{\RNum}[1]{\uppercase\expandafter{\romannumeral #1\relax}}

\newcommand{\AD}[1]{\textcolor{black}{#1}}

\graphicspath{{img/}}

\begin{document}
	
    \title{Optical Investigations of Coherence and Relaxation Dynamics of a Thulium-doped Yttrium Gallium Garnet Crystal at sub-Kelvin Temperatures for Optical Quantum Memory}

    \author{Antariksha Das}
    \altaffiliation{These authors contributed equally to this work. \newline Corresponding Author: wolfgang.tittel@unige.ch}
    \affiliation{QuTech and Kavli Institute of Nanoscience, Delft University of Technology, 2628 CJ Delft, The Netherlands}
    \affiliation{ICFO - Institut de Ciencies Fotoniques, The Barcelona Institute of Science and Technology, 08860 Castelldefels (Barcelona), Spain}

    \author{Mohsen Falamarzi Askarani$^*$}
    \affiliation{QuTech and Kavli Institute of Nanoscience, Delft University of Technology, 2628 CJ Delft, The Netherlands}
	
    \author{Jacob H. Davidson}
    \affiliation{QuTech and Kavli Institute of Nanoscience, Delft University of Technology, 2628 CJ Delft, The Netherlands}
	
    \author{Neil Sinclair}
    \affiliation{John A. Paulson School of Engineering and Applied Sciences, Harvard University, Cambridge, MA 02138, USA}
    \affiliation{Division of Physics, Mathematics, and Astronomy, and Alliance for Quantum Technologies, California Institute of Technology, Pasadena, California 91125, USA}

    \author{Joshua A. Slater}
    \affiliation{QuTech and Kavli Institute of Nanoscience, Delft University of Technology, 2628 CJ Delft, The Netherlands}
	
    \author{Sara Marzban}
    \affiliation{QuTech and Kavli Institute of Nanoscience, Delft University of Technology, 2628 CJ Delft, The Netherlands}

    \author{Daniel Oblak}
    \affiliation{Institute for Quantum Science and Technology, and Department of Physics \& Astronomy, University of Calgary, Calgary, Alberta, T2N 1N4, Canada}
	
    \author{Charles W. Thiel}
    \affiliation{Department of Physics, Montana State University, Bozeman, Montana 59717, USA}

    \author{Rufus L. Cone}
    \affiliation{Department of Physics, Montana State University, Bozeman, Montana 59717, USA}

    \author{Wolfgang Tittel}
    \affiliation{QuTech and Kavli Institute of Nanoscience, Delft University of Technology, 2628 CJ Delft, The Netherlands}
    \affiliation{Department of Applied Physics, University of Geneva, 1211 Geneva 4, Switzerland}
    \affiliation{Constructor University, 28759 Bremen, Germany}
	\date{\today}
	
\begin{abstract}
Rare-earth ion-doped crystals are of great interest for quantum memories, a central component in future quantum repeaters. To assess the promise of 1$\%$ Tm$^{3+}$-doped yttrium gallium garnet (Tm:YGG), we report measurements of optical coherence and energy-level lifetimes of its $^3$H$_6$ $\leftrightarrow$ $^3$H$_4$ transition at a temperature of around 500 mK and various magnetic fields. Using spectral hole burning, we find hyperfine ground-level (Zeeman level) lifetimes of several minutes at magnetic fields of less than 1000 G. We also measure coherence time exceeding one millisecond using two-pulse photon echoes. Three-pulse photon echo and spectral hole burning measurements reveal that due to spectral diffusion, the effective coherence time reduces to a few $\mu$s over a timescale of around two hundred seconds. Finally, temporal and frequency-multiplexed storage of optical pulses using the atomic frequency comb protocol is demonstrated. Our results suggest Tm:YGG to be promising for multiplexed photonic quantum memory for quantum repeaters.

\end{abstract}


\maketitle

\section{Introduction}

Cryogenically cooled rare-earth-ion-doped crystals (REIC) \cite{REI-Book, REI1, REI2, REI3} are promising candidates for optical and quantum memory applications \cite{QM1}, as demonstrated in a series of recent publications, including storage of quantum states over long times \cite{20ms_storage, 1hr_storage}, with high efficiencies \cite{Efficient_storage_1, Efficient_storage_2, Efficient_storage_3, cold_atoms1}, in multiple modes \cite{saglamyurek2016multiplexed, yang2018multiplexed, pu2017multiplexed}, and of entangled memories \cite{heralded_storage_mohsen, rakonjac2021entanglement}. Quantum memories have been realized using various physical systems such as hot and cold atomic vapor \cite{hot_Rb, Efficient_storage_2, Efficient_storage_3, Rb_Raman, cold_atoms1, cold_atoms2}, single-trapped atoms, and diamond colour centers \cite{NV_quantum_memory_1, NV_quantum_memory_2, NV_quantum_memory_3, NV_quantum_memory_4, NV_quantum_memory_5}, with each system having its own advantages. The interest in using rare-earth-doped solids \cite{REI1, REI2, REI3} is due to their long coherence lifetimes \cite{4msT2Er, 6msT2Eu, Das, 1hr_storage} observed for both optical and spin transitions, and their large optical inhomogeneous broadening in conjunction with long population lifetimes of hyperfine transitions. For example, these properties are well suited to implement storage protocols based on atomic frequency combs (AFCs), which enables storage with a large time-bandwidth product and allows multiplexing. Quantifying the time-dependent spectroscopic properties of REICs at temperatures near absolute zero is both of fundamental interest and a prerequisite for their use as quantum memories.

Thulium-doped crystals \cite{Charles2012Tm:LN, louchet2008optical, Charles1, Charles2} are desirable candidates for spectrally-multiplexed quantum memories with fixed storage time \cite{Multiplexing}, an approach that relies on the two-level atomic frequency comb (AFC) protocol \cite{AFC, QM2_echo}. These crystals possess the simplest energy level structure that allows for spectral tailoring as needed to create AFCs: both the ground and excited states split under the application of a magnetic field into two long-lived Zeeman levels, which allows for persistent spectral hole burning \cite{PersistentAFC}. Furthermore, the optical coherence time, which determines the maximum storage time, is generally long, e.g. 119 $\mu$s in Tm:YAG \cite{REI-Book, macfarlane1993photon}.

In this work, we investigate the spectroscopic properties of a 1$\%$ thulium-doped ytterbium gallium garnet (Tm$^{3+}$:Y$_3$Ga$_5$O$_{12}$ or Tm:YGG) crystal at temperatures as low as 500 mK and various magnetic fields. We measure the optical coherence times T$_2$ and lifetimes of all energy levels that are relevant for the AFC protocol by means of two-pulse photon echo (2PPE) and time-resolved spectral hole burning (SHB) techniques. We find a sub-kHz wide homogeneous linewidth $\Gamma_{h}$ (with $\Gamma_{h}$=1/$\pi T_2$) - the third narrowest reported for any optical transition after Erbium and Europium \cite{4msT2Er,6msT2Eu} - as well as hyperfine levels (Zeeman levels) within the electronic ground state with up to 300 sec lifetime. Furthermore, we observe and investigate the magnetic field-induced broadening of spectral holes over time, which we explain qualitatively in the context of spectral diffusion. Pathways towards improved long-term coherence are suggested. The exceptional homogeneous linewidth together with a 56 GHz inhomogeneously broadened absorption profile \cite{Charles2} and long-lived atomic levels makes Tm:YGG a promising candidate for multimode quantum memories that enable spectrally multiplexed quantum repeaters.

Our paper is organized as follows. Sections \RNum{2} and \RNum{3} discuss the properties of the Tm$^{3+}$:YGG crystal used in our experiment and describe the experimental setup.  In Section \RNum{4}, measurements of Zeeman-level lifetimes and optical coherence are presented. Then, in Section \RNum{5}, we study spectral diffusion in the presence of magnetic fields over different timescales. Toward this end, the results of three-pulse photon echo experiments and spectral hole burning measurements are presented, and plausible underlying decoherence mechanisms are proposed and discussed. In Section \RNum{6}, we demonstrate microsecond-long storage of 10 subsequent laser pulses and simultaneous storage of 3 different spectral modes using the AFC echo scheme. We end the paper with a conclusion and an outlook.

\section{Tm:YGG Material Properties}

For our measurements, we use a 25 mm $\times$ 5 mm $\times$ 5 mm ($a \times b \times c$) single crystal of  1\% Tm$^{3+}$:Y$_3$Ga$_5$O$_{12}$ from Teledyne FLIR Scientific Materials (Bozeman, MT). Y$_3$Ga$_5$O$_{12}$ (YGG) is a cubic crystal in which all naturally occurring isotopes of the host ions Y and Ga feature nuclear spin: I = $\frac{1}{2}$ for $^{89}$Y and I = $\frac{3}{2}$ for $^{69}$Ga and $^{71}$Ga \cite{Persson20142}. \AD{Yttrium has a natural abundance of 100$\%$ with a free-nucleus gyromagnetic ratio of 2.1 MHz/T. For gallium, the natural abundance of $^{69}$Ga is 60$\%$ and a free-nucleus gyromagnetic ratio of 10.2 MHz/T, while the $^{71}$Ga isotope has a natural abundance of 40$\%$  with a free-nucleus gyromagnetic ratios of 13 MHz/T \cite{lee1967table}.} Tm$^{3+}$ ions substitute Y$^{3+}$ without charge compensation in six crystallographically equivalent but orientationally inequivalent sites of D$_2$ point symmetry. The transition between the lowest crystal-field levels of the electronic ground state, $^3$H$_6$, and the optically excited state, $^3$H$_4$, occurs at a wavelength of 795.325 nm in vacuum. A simplified energy-level diagram of Tm$^{3+}$:YGG is shown in the inset of Fig. \ref{fig:setup}. Tm$^{3+}$ is a non-Kramers ion with a [Xe]4$f^{12}$ electron configuration, the energy levels are electronic singlets, and the angular momentum is quenched by the crystal field, resulting in no first-order hyperfine interaction. Due to the I= $\frac{1}{2}$ nuclear spin of Tm$^{3+}$, the quadrupole interaction is zero and the second-order magnetic hyperfine interaction vanishes in the absence of a magnetic field. When an external magnetic field is applied, nuclear Zeeman and electronic Zeeman interactions combine with a second-order hyperfine interaction to produce an ``enhanced" effective nuclear magnetism that splits the ground and excited states into pairs (m$_I$ = $\pm \frac{1}{2}$) of sublevels. We refer to these levels as Zeeman levels, but the splitting mechanism includes all interactions mentioned above. The Zeeman level structure is hidden within the $\sim$ 56 GHz inhomogeneous broadening of the $^3$H$_6$$\leftrightarrow$$^3$H$_4$ optical transition.

\section{Experimental details}

\begin{figure}[h!]
\centering
\includegraphics[width = 1\linewidth]{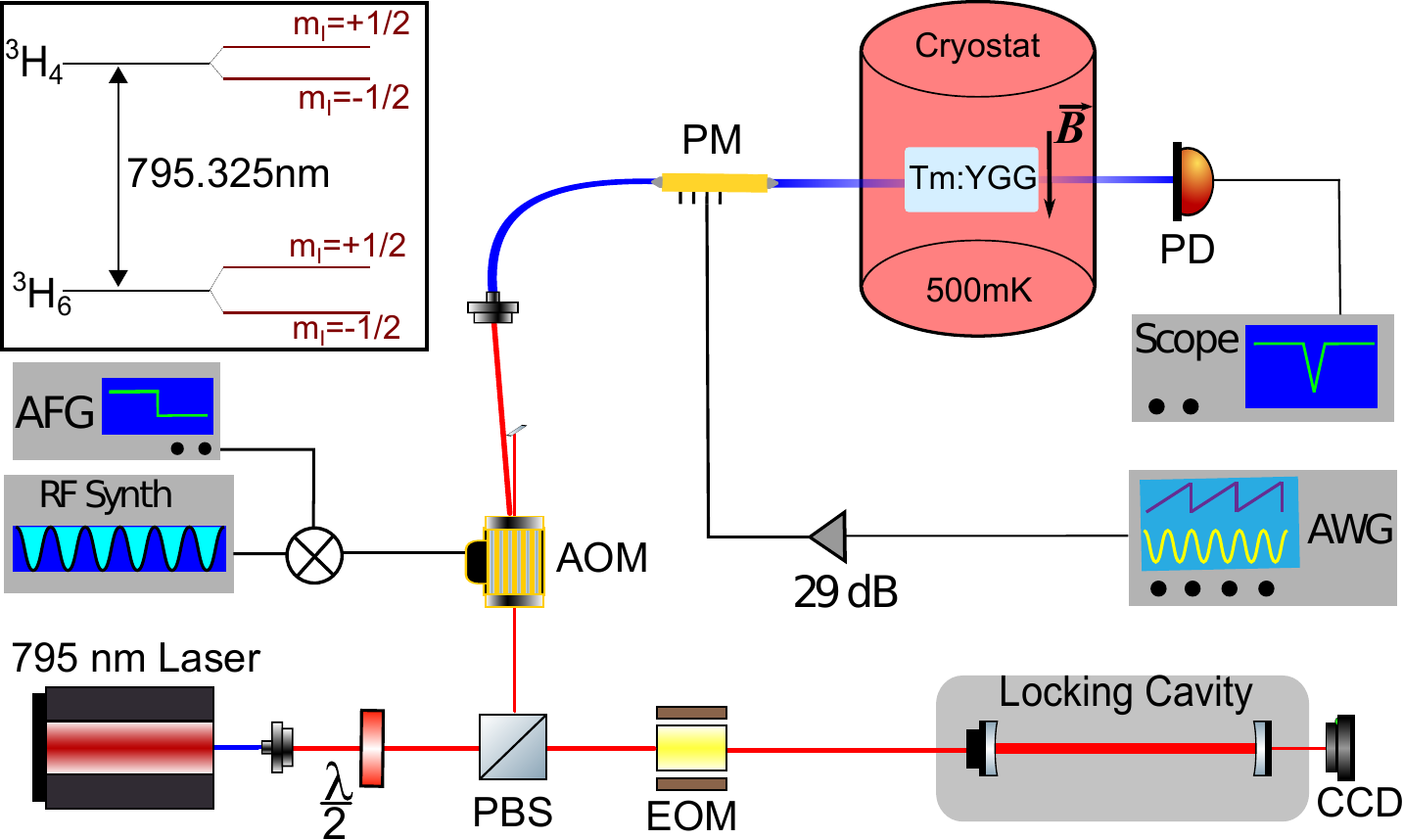}
\caption{\textbf{A schematic diagram of the experimental setup:} AFG: arbitrary function generator; RF Synth: radio frequency synthesizer; AOM: acousto-optic modulator; AWG: arbitrary waveform generator; PM: phase modulator; EOM: electro-optic modulator; PD: photo detector; CCD: charge-coupled device camera; \textit{\textbf{B}}: Magnetic field. Inset: Simplified energy level diagram of the $^3H_6$ $\rightarrow $ $^3H_4$ optical transition of $\mathrm {Tm^{3+}}$ in $\mathrm {Y_3Ga_5O_{12}}$(YGG) crystal. Only the lowest crystal field level of each electronic manifold is shown.}
\label{fig:setup}
\end{figure}
A schematic of our setup is presented in Fig. \ref{fig:setup}. For our measurements the crystal is mounted on the coldest stage of an adiabatic demagnetization refrigeration (ADR)-based cryostat, reaching a temperature below 600 mK. The light propagates along the 25 mm long $\langle110\rangle$ axis of the crystal, and its polarization is linear at the crystal input. The polarization evolves inside the crystal due to birefringence stemming from residual crystal strain. A magnetic field is applied parallel to the $\langle 111 \rangle$-direction using a superconducting solenoid. The field strength is detected using a Hall sensor mounted directly above the crystal.

The optical pulse sequence for our measurements is obtained from a continuous-wave external-cavity Toptica DLPro tunable diode laser emitting at 795.32 nm (vacuum) to address the $^3$H$_6$ $\rightarrow$ $^3$H$_4$ transition of Tm$^{3+}$ ions in the crystal. A programmable pulseblaster controls the timing sequence. The light is gated and shaped with a single-pass free-space acoustic-optic modulator (AOM). The 1st order diffracted light beam from the AOM is sent through a  series of optical components (waveplates and polarizing beam-splitter) and a phase modulator (PM) where the frequency of the light is shifted through a serrodying signal \cite{serrodyne1} created by an arbitrary waveform generator (AWG). Optical transmission through the crystal is detected using an amplified silicon photodetector and a digital oscilloscope.

A small fraction of light from the diode laser is used to frequency lock the laser to a high-finesse cavity using the Pound-Drever-Hall method \cite{black2001introduction, singh2020stabilization}. For improved stability, the frequency stabilization setup is kept on a separate optical breadboard that is mechanically isolated from the main optical setup. After passing through an electro-optic modulator (EOM) that is used to create frequency sidebands, the light is sent through a series of optical components for proper coupling into the temperature-stabilized high-finesse optical cavity through an optical fiber that is kept under vacuum. The reflected signal is measured using a photodiode, creating a feedback loop via a PID controller to suppress laser frequency fluctuations. This allows us to obtain a narrow laser linewidth of a few tens of kilohertz.

\section{Spectroscopic results}

\subsection{Spectral hole burning measurements}

\begin{figure*}[ttt]
\centering
\includegraphics[width = 1\linewidth]{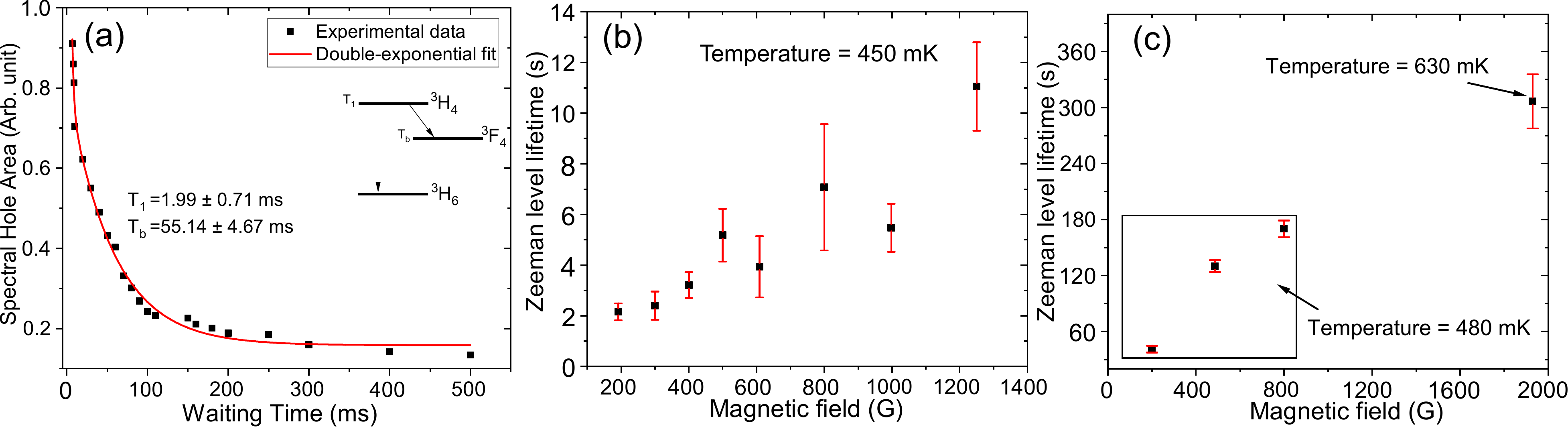}
\caption{\textbf{Energy-level lifetime measurements:} (a) A measurement of population decay, probed using spectral hole areas that allow extracting the $^3$H$_4$ excited and $^3$F$_4$  bottleneck level lifetimes. The plots in (b) and (c) show the extracted lifetimes of the `short' and `long' decay of the Zeeman level for varying magnetic fields, associated with the two magnetically inequivalent subgroups of Tm$^{3+}$ ions.}
\label{fig:Zeeman lifetime}
\end{figure*}

To determine the lifetime of different energy levels and the Zeeman levels of Tm$^{3+}$:YGG, we employ a widely used spectroscopic technique known as time-resolved spectral hole burning (SHB). We set up a hole-burning sequence composed of the following steps. First, the ions in the crystal are optically pumped from the ground to the excited state using a 50 ms-long monochromatic laser pulse. After a waiting time that varies between a few tens of microseconds and a few hundred milliseconds, the hole is read out with an attenuated 8 ms long pulse that is frequency chirped over 2 MHz. By measuring the area of the spectral hole for varying waiting times, the lifetimes of the various levels that lie between the excited state and the ground state can be extracted. An example of a decay curve is shown in Fig. \ref{fig:Zeeman lifetime}a. Fitting this spectral hole decay using a double-exponential function reveals a $^3$H$_4$ excited state lifetime T$_1$ = $1.99\pm 0.71$ ms and a $^3$F$_4$ bottleneck level lifetime T$_b$ = $55.14\pm4.67$ ms, which agrees with previous results \cite{Charles1}.

Next, to measure the ground state $^3$H$_6$ Zeeman level lifetime, a magnetic field is applied along the $\langle111\rangle$ axis of the crystal with the sample at a temperature of T = 500 mK. Again, by measuring the persistent spectral hole decay as a function of waiting time for different magnetic fields, varying from a few hundred Gauss to a few thousand Gauss, we obtain the ground state Zeeman level lifetime as a function of different magnetic field strengths, which is shown in Fig. \ref{fig:Zeeman lifetime}. Due to the $D_2$ point symmetry of the Tm$^{3+}$ ions in the YGG lattice, there are six subgroups of magnetically inequivalent Tm$^{3+}$ ions that have different local site orientations. Different subgroups of ions can be selectively addressed by choosing the orientation of the magnetic field as well as the propagation direction and the optical polarization of the light \cite{Tm_polarization_PRB, de2006experimental}. For our experiment, the magnetic field is oriented along the $\langle111\rangle$ direction and the light is propagating along the $\langle110\rangle$ direction. In this configuration, two of the six subgroups of magnetically inequivalent Tm$^{3+}$ ions interact with the light \cite{Veissier2016, Davidson}. Thus, we observe two different sets of Zeeman level lifetimes \cite{Charles1, sinclair2017properties} as indicated in Fig. \ref{fig:Zeeman lifetime}b. and Fig. \ref{fig:Zeeman lifetime}c. respectively. It's worth noting that in our experiment, we obtain a combined optical response from these two sites. However, it is possible to selectively choose either of the sites by appropriately aligning the polarization of the light \cite{das_quadratic}. The persistence of Zeeman levels enables the creation of long-lived spectral holes, which, in turn, allow tailoring the absorption profile of the crystal and to prepare persistent atomic frequency combs (AFC) \cite{AFC}, a well-known quantum memory protocol used to store single photons in an inhomogeneously broadened material.

\subsection{Optical coherence time measurements}

To study the optical coherence properties in Tm$^{3+}$:YGG, we employ two-pulse photon echo (2PPE) spectroscopy \cite{allen1987optical}. In a 2PPE sequence, two excitation pulses separated by a waiting time $t_{12}$ are sent into an inhomogeneously broadened ensemble of resonant Tm$^{3+}$ ions. They prepare a coherent superposition of the ground and excited electronic states. This gives rise to a coherent burst of radiation---a photon echo---at time $t_{12}$ after the second pulse. The variation of the echo intensity as a function of $t_{12}$ can be written as: 
\begin{equation}
    I(t_{12}) = I_0 e^{-4\pi\Gamma_{h}t_{12}}
    \label{mims1}
\end{equation}
where $I_0$ is the initial echo intensity at $t_{12}$=0 and $\Gamma_h$ is the homogeneous linewidth, which is inversely proportional to the coherence time $T_2$ :
\begin{equation}
    \Gamma_h = \frac{1}{\pi T_2}
    \label{mims2}
\end{equation}
In order to extract the effective homogeneous linewidth as a function of the magnetic field, we vary the magnetic field from 100 G to 1000 G and measure two-pulse photon echo decays at a temperature of 500 mK and a wavelength of 795.32 nm. We fit all measured photon echo decays using the exponential function described in Eq. \ref{mims1}. 

\begin{figure}[h!]
\centering
\includegraphics[width = 0.9\linewidth]{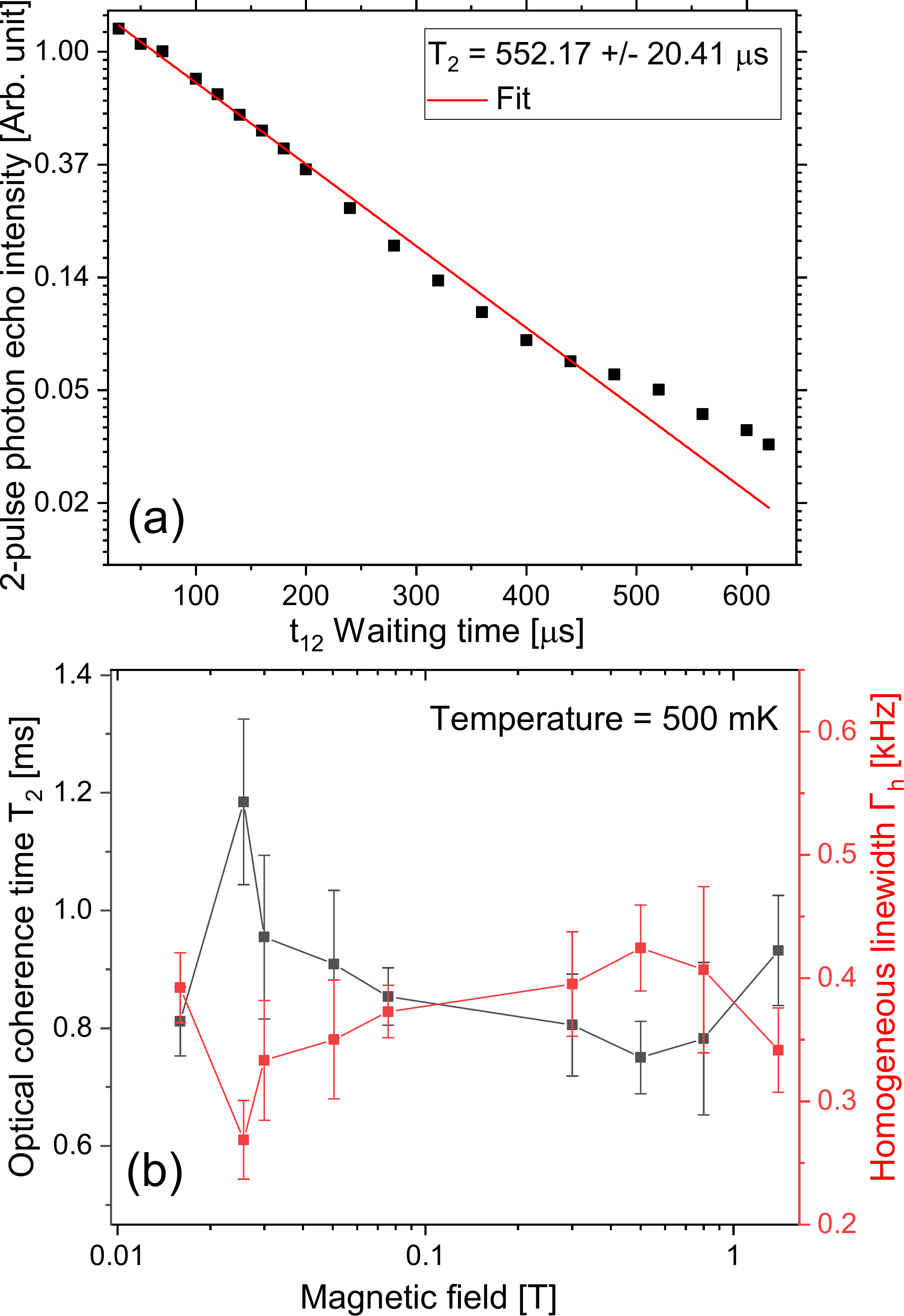}
\caption{\textbf{Optical coherence time measurements:} (a) Two-pulse photon echo decay as a function of waiting time t$_{12}$  between the $\frac{\pi}{2}$ and the $\pi$ pulses at zero magnetic field. (b) Optical coherence time T$_2$, and corresponding homogeneous linewidth $\Gamma_{h}$, as a function of magnetic field strength. All data points are connected by solid lines as a guide to the eye.}
\label{fig:T2vsB}
\end{figure}

The magnetic field dependence of the coherence time $T_2$ and of the homogeneous linewidth $\Gamma_h$ is presented in Fig. \ref{fig:T2vsB}b. They reflect the dominant magnetic field-dependent decoherence processes \cite{Charles2012Tm:LN} which limit the performance of the crystal as a quantum memory. We find that the introduction of a few hundred Gauss magnetic fields improves the coherence time $T_2$ from a zero-field value of 552 $\mu$s (see Fig. \ref{fig:T2vsB}a) to a maximum of around 1.1 ms, corresponding to a minimum homogeneous linewidth $\Gamma_h$ of around 0.26 kHz at around 200 G \cite{Das}. For larger fields, the coherence time decreases to around 0.8 ms and remains approximately constant up to 1 Tesla (see Fig. \ref{fig:T2vsB}b).

\section{Spectral diffusion}

Spectral diffusion results in broadening of the homogeneous linewidth $\Gamma_h$ as a function of time because each Tm$^{3+}$ ion experiences a slightly different dynamic crystalline environment. Spectral diffusion is expected due to the presence of gallium and yttrium in the YGG lattice, both of which have nuclear spins and may couple to Tm$^{3+}$ ions. The application of a magnetic field generally reduces the impact of spectral diffusion by inhibiting nuclear spin flips \cite{REI-Book, Bottger}. Two well-known physical mechanisms that can be accountable for 
spin flips are phonons (spin-lattice relaxation) \cite{Thomas_decoherence, Mohsen_decoherence}, and spin-spin relaxation through magnetic dipole-dipole interaction \cite{Bottger}, which causes pairs of anti-parallel spins to flip simultaneously (spin flip-flops). These correlated spin flips can randomize the local spin orientations \cite{Bottger, Decoh1}. \AD{For the range of magnetic fields and temperatures examined in our work, the nuclear spin splittings are comparable to homogeneous linewidths and nuclear quadrupole splitting, resulting in homo and heteronuclear transition resonances \cite{Davidson}. The magnitude of the resulting magnetic field fluctuations at the Tm sites is sufficient to cause up to a MHz of decoherence, depending on the timeframe of the fluctuations.} This fluctuating magnetic field within the YGG crystal can be due to dynamic interactions between host nuclear spins ($^{69}$Ga, $^{71}$Ga, Y$^{3+}$) and paramagnetic impurities. In order to characterize the spectral diffusion beyond the 2PPE timescales in the presence of a magnetic field, we employ three-pulse photon echo (3PPE) spectroscopy and spectral hole burning.


\subsection{Three-pulse photon echo (3PPE) measurements}

\begin{figure*}[ttt]
\centering
\includegraphics[width = 0.9\linewidth]{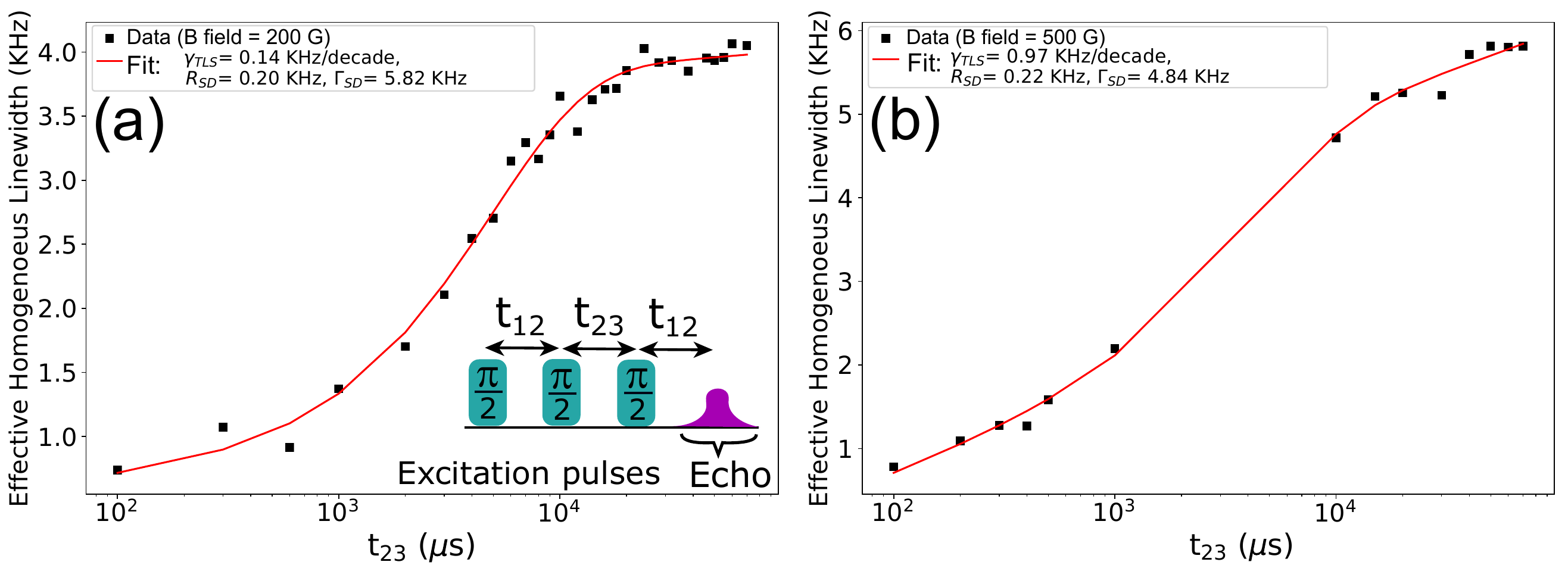}
\caption{\textbf{Time-dependence of optical coherence time, evaluated using three-pulse photon echoes:} Effective homogeneous linewidth as a function of time delay $t_{23}$ at 500 mK temperature for an applied magnetic field of (a) 200 G and (b) 500 G. The solid lines represent the fits. Inset: Three-pulse photon echo sequence. The delay $t_{12}$ is between the first and second pulse, and $t_{23}$ is between the second and third pulse.}
\label{fig:3PPE}
\end{figure*}

The effective homogeneous linewidth and the time evolution of spectral diffusion-induced decoherence can be extracted from three-pulse photon echo measurements. In a 3PPE sequence, the first two excitation pulses, separated by a waiting time $t_{12}$, are sent into an inhomogeneously broadened ensemble of absorbers to create a frequency-dependent periodic modulation of the population in the ground and excited states. Then, after a time delay $t_{23}$, a third pulse is applied, resulting in the emission of an echo a time $t_{12}$ after the third pulse. To investigate spectral diffusion of Tm$^{3+}$ ions at timescales up to 100 ms, we performed 3PPE measurements at a temperature T = 500 mK and a magnetic field of a few hundred Gauss. For our measurements, the separation time $t_{12}$ between the first two pulses is held constant at 60 $\mu$s, and the echo intensity is measured as a function of time delay $t_{23}$ between the second and third pulse, with $t_{23}$ varying between 100 $\mu$s and 100 ms. The echo intensity is given by \cite{Mohsen_decoherence, neil_tmln, Bottger}
\begin{equation}
\begin{aligned}
I\left(t_{23}\right)=I_0 I_{\text {pop}}^2(t_{23}) e^{-4 t_{12} \pi \Gamma_h\left(t_{23}\right)}
\end{aligned}
\label{SD}
\end{equation}
where \textit{I}$_0$ is a scaling coefficient and \textit{I}$_{\mathrm{pop}}$ \AD{captures the effect of the reduction of contrast of the atomic grating caused by population decay, which, consequently, reduces the echo intensity}. For Tm$^{3+}$:YGG, $I_{\mathrm{pop}}\left(t_{23}\right) \approx C_1 e^{-t_{23} / T_1}+C_B e^{-t_{23} / T_B}+C_Z e^{-t_{23} / T_Z}$, where $C_1$, $C_B$, $C_Z$ are constants. T$_1$, T$_B$, and T$_Z$ are the excited state lifetime ($^3$H$_4$), the bottleneck level lifetime ($^3$F$_4$), and the Zeeman level lifetime ($^3$H$_6$ (m$_I$ = $\frac{1}{2}$)), measured in Section \RNum{4} A. $\Gamma_{\textrm{eff}} (t_{23})$ is the time-dependent ``effective" homogeneous linewidth. It captures all diffusion processes that influence the rare-earth spins caused by magnetic dipole-dipole interactions. Following \cite{Bottger, Mohsen_decoherence, neil_tmln}, the functional form of $\Gamma_{\textrm{eff}} (t_{23})$ can be written as 
\begin{multline}
\Gamma_{\textrm{eff}} (t_{23})
= \Gamma_0 +\gamma_{TLS} \textrm{ log}(\frac{t_{23}}{t_0}) \\
+ \frac{1}{2} \Gamma_{SD} (R_{SD} t_{12} + 1 - e^{-R_{SD} t_{23}})
\label{SD1}
\end{multline}
where $\Gamma_0$ is the homogeneous linewidth at the minimum measurement timescale ${t_0}$ (160 $\mu$s in our experiment), $\Gamma_{SD}$ is the maximum broadening of the homogeneous linewidth (or the spectral diffusion linewidth), and $R_{SD}$ describes the characteristic diffusion rate of linewidth broadening. The values of these parameters are determined by the details of the diffusion mechanisms \cite{Charles2}. We also consider spectral diffusion due to thermally activated low-energy dynamic structural fluctuations, often described as two-level systems (TLS) \cite{TLS1, TLS2}, with $\gamma_{TLS}$ being the TLS mode coupling coefficient.

To characterize the effects of spectral diffusion, we fit the measured echo decays using Eq. \ref{SD} and extract the effective homogeneous linewidth, which is plotted in Fig. \ref{fig:3PPE} for T = 500 mK and for two different magnetic fields as a function of the time-delay $t_{23}$. Fitting each curve to Eq. \ref{SD1} yields a homogeneous linewidth $\Gamma_0$ of a few hundred Hz and that the spectral diffusion saturates at a maximum value of a few kHz. Based on the magnitude of the spectral diffusion parameters, $\Gamma_{SD}$ and R$_{SD}$ ($\Gamma_{SD}$ = 5.82$\pm$0.52 kHz, $R_{SD}$ = 0.20$\pm$0.01 kHz at 200 G and $\Gamma_{SD}$ = 4.84$\pm$0.72 kHz, $R_{SD}$ = 0.22$\pm$0.04 kHz at 500 G), it is likely that the dominant source of spectral diffusion stems from nuclear spin flips of neighboring gallium in the host lattice. We also find that the contribution of low-energy TLS modes to linewidth broadening is not very pronounced at this temperature and that the spectral diffusion parameters in the assessed temporal regime show little magnetic field dependence.  Additional measurements across a broad range of parameters are necessary to conclusively confirm the existence of broadening attributed to TLS.


\subsection{Long-term spectral diffusion: Magnetic-field-dependent spectral hole broadening}

\begin{figure*}[ttt]
\centering
\includegraphics[width = 1\linewidth]{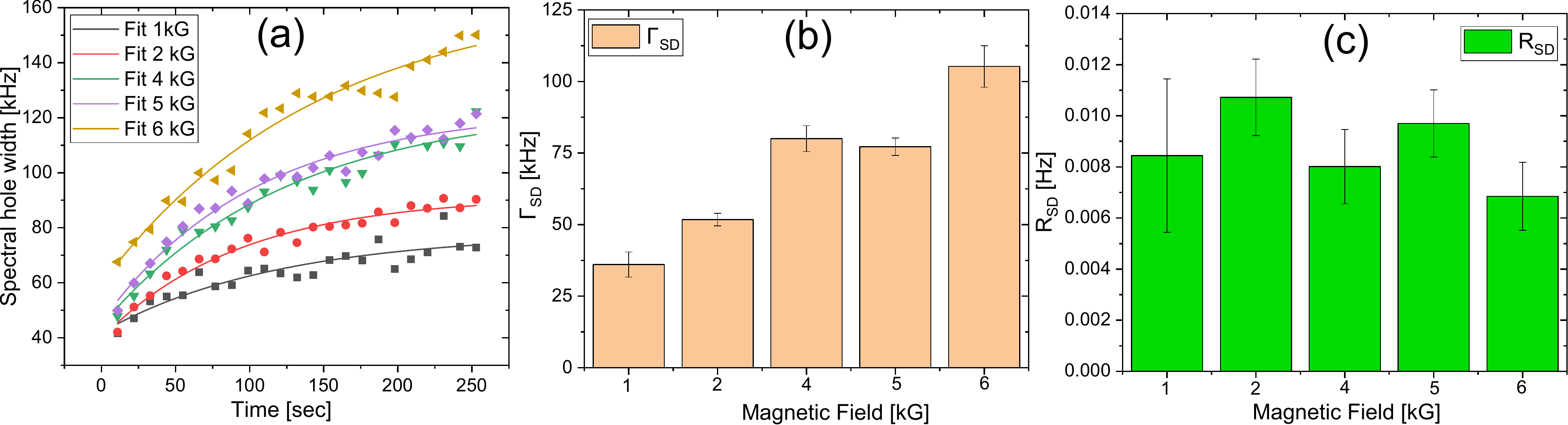}
\caption{\textbf{Spectral diffusion:} (a) Spectral hole widths as a function of time at 600 mK temperature and magnetic fields between 1 kG and 6 kG. The solid lines are the best fits. (b) Variation of the extracted spectral diffusion linewidth $\Gamma_{SD}$ as a function of the magnetic field strength. (c) Variation of the linewidth broadening rate R$_{SD}$ as a function of the magnetic field strength.}
\label{fig:hole_broadening}
\end{figure*}
In the presence of a magnetic field, spectral diffusion is known to occur over timescales on the order of Zeeman-level lifetimes, which, in our crystal, are many seconds. Since many optical signal processing applications rely on similarly long-lived spectral features \cite{REI-Book, REI2}, we investigate spectral diffusion on such timescales. However, since the characteristic timescales for three-pulse photon echo measurements are much smaller, we instead observe the broadening of persistent spectral holes, created by changing the Zeeman level population out of equilibrium. We operate at a temperature of 600 mK and magnetic fields between 1 and 6 kG. The burning duration and the waiting time are adjusted to maximize the initial hole depth. Once the hole burning is completed, the spectral hole is read out by a weak probe pulse. Then, we determine the spectral hole widths for a series of different magnetic fields. And finally, we observe how the spectral hole broadens as a function of time, see Fig. \ref{fig:hole_broadening}a. Assuming no power broadening, the spectral hole width is proportional to the homogeneous linewidth ($\Gamma_{\mathrm{spectral-hole}}$ = 2 $\Gamma_{\mathrm{h}}$), and the observed behavior can be described by \cite{Bottger}
\begin{align} \label{sudden Jump}
\Gamma_{\mathrm{spectral-hole}} (t) \propto \Gamma_{0} + \frac{1}{2}\Gamma_{SD}(1- e^{-R_{SD}t})
\end{align}

Upon fitting the spectral hole broadening depicted in Fig. \ref{fig:hole_broadening}a with Eq. \ref{sudden Jump}, we find that the spectral diffusion linewidth $\Gamma_{SD}$ increases from approximately ~36 kHz to about ~105 kHz as the applied magnetic field strength is increased from 1 kG to 6 kG, as shown in Fig. \ref{fig:hole_broadening}b. Furthermore, it is worth noting that, regardless of the magnetic field strength, the broadening exhibits a nearly constant rate of approximately 0.01 Hz, as highlighted in Fig. \ref{fig:hole_broadening}c. It is important to emphasize that, within the time scales, magnetic field, and temperature under consideration, the relaxation rate R$_{SD}$ is significantly slower than the population decay rate \textit{I}$_{pop}$ in Eq. \ref{SD} and is therefore not observable using 3PPE measurements.

The relationship governing the spectral diffusion linewidth, $\Gamma_{SD}$ and the applied magnetic field can be expressed as follows:\cite{Bottger},
\begin{align} \label{sech}
\Gamma_{SD} (B,T) = \Gamma_{max}(B) \sech^2(g\mu_{b}B/2k_{B}T)
\end{align}
where, $g$ represents the g-factor associated with the spins present in the crystal lattice., $\mu_{B}$ denotes the Bohr magneton, $B$ is the external magnetic field, $k_{B}$ is the Boltzmann constant, $T$ corresponds to the temperature under consideration, and $\Gamma_{max}(B)$ represents the maximum frequency broadening of the optical transition resulting from magnetic dipole-dipole interactions, \AD{which also depends on the magnetic field $B$}. 

In Tm$^{3+}$:YGG, the g-factors of thulium ($0.01$), gallium ($7.14 \times 10^{-4}$ and $9.2\times 10^{-4}$ for the two isotopes), and yttrium ($1.43\times 10^{-5}$) spins are notably small \cite{WebElements}. \AD{We must consider the magnitude of the $\sech^2$ term compared to that of $\Gamma_{max} (B)$ in Eq. \ref{sech}}. Moreover, for the range of temperatures and magnetic fields examined in our work, the ratio of the thermal distribution of atomic population in Zeeman levels remains relatively constant. Consequently, the hyperbolic secant term on the R.H.S of Eq. \ref{sech} remains effectively constant. As a result, the spectral diffusion predominantly depends on the magnitude of the magnetic dipole moment induced in the Tm$^{3+}$ ions by the external magnetic field, which is directly proportional to $\Gamma_{max}(B)$. 

Below, we outline potential underlying physical processes that could lead to variations of the magnetic field at the Tm$^{3+}$ ion locations, and thus, contribute to the spectral diffusion:

\begin{itemize}
  \item Ion-ion coupling: Spectral diffusion due to Tm$^{3+}$-Tm$^{3+}$ interaction can play a role in the observed broadening of the linewidth. The low Tm$^{3+}$ ion concentration of 1$\%$ in this YGG crystal signifies that the average distance between thulium ions is unlikely to cause broadening of the observed magnitude but an excitation-induced interaction between Tm$^{3+}$ ions \cite{Thiel_2014} or strain-mediated Tm$^{3+}$-Tm$^{3+}$ interaction \cite{louchet2023strain} may cause this to happen. \AD{This is similar to the situation observed in Tm:YAG, which has the same site symmetry and also has negligible magnetic or electric dipole-dipole interactions, but still exhibits very strong instantaneous spectral diffusion that is comparable in magnitude to the magnetic interaction strength.}
  \item Phonon-induced spin-flips: As the magnetic field increases, the phonon density of states increases quadratically, leading to an increase in the probability of interaction between the phononic modes of the crystalline lattice and thulium ions \cite{Thomas_decoherence}.
  \item Quadratic Zeeman effect: \AD{The quadratic Zeeman effect arises from mixing of the crystal-field levels due to the applied magnetic field. This causes shifts in the energy levels in both the ground and excited states, leading to a shift of the optical transition frequency that is proportional to \textit{B}$^2$ and strongly orientation and site dependent.} The observed spectral broadening can also originate from the interaction between thulium ions and neighboring host ions. The neighboring nuclear spins surrounding the Tm$^{3+}$ ion are yttrium Y$^{3+}$ and two isotopes of gallium, $^{69}$Ga and $^{71}$Ga. Since Y$^{3+}$ nuclear spins are weakly magnetic, it is plausible that there exists a highly concentrated spin bath comprised of gallium nuclear spins, each possessing a moderate nuclear magnetic moment, which can cause variations in the magnetic field experienced by the thulium ions through the quadratic Zeeman effect \cite{Veissier2016}. \AD{We have extensively investigated quadratic Zeeman spectral diffusion in Tm:YGG in our prior work \cite{das_quadratic}.}
 \item Fluctuating external magnetic field: An additional contribution may be a noisy current supply for the superconducting solenoid magnet that would cause magnetic field variations of $\Gamma_{SD}$ or stray radio-frequency fields that could drive nuclear spin flips.
\end{itemize} 

\section{Multiplexed optical storage}

The spectroscopic investigations show that Tm:YGG is suitable for quantum memory applications. To verify this conjecture we demonstrate spectrally and temporally multimode atomic frequency comb (AFC)-based storage of laser pulses. The two-level atomic frequency comb (AFC) protocol \cite{AFC} is well known and well established for storage of quantum light as well as classical laser pulses in REI-doped solids \cite{Das, Multiplexing, 1hr_storage}. It relies on shaping the inhomogeneously broadened absorption profile of an ensemble of absorbers into a series of equally spaced teeth by means of persistent spectral hole burning with peak separation $\Delta$. The absorption of a photon by such a comb yields a collective atomic excitation that can be described by a so-called Dicke state:
\begin{equation}
    |\Psi\rangle=\frac{1}{\sqrt{N}} \sum_{j=1}^{N} C_{j} e^{-i 2 \pi \delta_{j} t} e^{i k z_{j}}\left|g_{1}, \ldots, e_{j}, \ldots, g_{N}\right\rangle
\end{equation}
where $\ket{e_j}$ represents the $jth$ atom being in the excited state, $\delta_j$ is the detuning of its atomic transition frequency with respect to the central frequency of the absorbed photon, $z_j$ is its position measured along the propagation direction of the light, $k$ the wavevector, and $C_j$ its absorption probability amplitude. The collective excitation described by the Dicke state subsequently dephases but, due to the periodicity of the comb, rephasing of the atomic excitations will occur at time 1/$\Delta$. This results in a photon-echo-like re-emission of the stored light. An important figure of merit of an AFC is its finesse, which is defined as the ratio between the peak separation and the FWHM of the absorption peaks, F = $\frac{\Delta}{\gamma}$.

Unlike other storage protocols, the multimode capacity of the AFC memory protocol does not depend on the optical depth of the storage medium, making this protocol a natural choice for multiplexed quantum memories \cite{Multiplexing, saglamyurek2016multiplexed, das2022long}. Its large temporal multimode capacity can readily be combined with multiplexing in frequency and space. In the time domain, the number of temporal modes that can be stored using the AFC scheme is proportional to the number of comb teeth, which depends on the total bandwidth and the periodicity $\Delta$. In the frequency domain, the number of spectral modes depends on the bandwidth per spectral channel and the total absorption bandwidth of the rare-earth crystal. In the following subsection, we demonstrate a multimode AFC memory in both the temporal and spectral domains. While we perform the experiments using classical optical pulses, it is known that the AFC protocol also allows storing quantum states of light such as qubits with high fidelity \cite{Entanglement_and_nonlocality}.

\begin{figure*}[ttt]
\centering
\includegraphics[width = 1\linewidth]{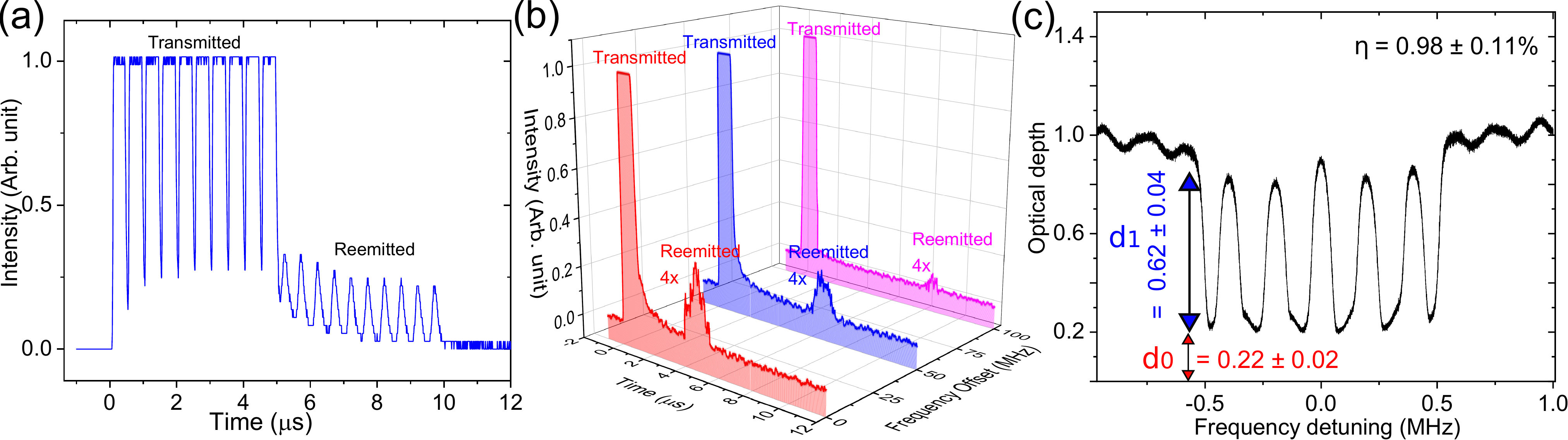}
\caption{\textbf{ Multiplexed storage of optical pulses using the AFC protocol:} (a) Storage of 10 temporal modes of 200 ns duration during 5~$\mu$s. The re-emitted pulse train is observed in the temporal window extending from 5~$\mu$s to 10~$\mu$s. (b) Three frequency-multiplexed AFCs are used to simultaneously store optical pulses of 1~$\mu$s duration in 3 spectral modes. Retrieval of each mode is shown in different colors. The spectral channels feature 4~$\mu$s, 6~$\mu$s, and 8~$\mu$s storage time, respectively, and their relative detuning is 50 MHz. (c) The absorption spectrum of a 1 MHz-wide AFC of finesse 2 tailored for 5~$\mu$s storage time.}
\label{fig:multiplexing}
\end{figure*}

\subsection{Simultaneous storage of subsequent temporal modes}

To demonstrate multimode AFC storage in time, we prepare a 10 MHz broad AFC with finesse F = 2 by optically pumping Tm$^{3+}$ ions to long-lived Zeeman levels. To avoid spontaneous emission noise due to population decay from the $^3$H$_4$ excited-state level, we wait 10 ms (five times the 2 ms lifetime of this level). Then, we create a sequence of 10 subsequent pulses of 200 ns duration and 100 ns spacing and send them into the memory. An attenuated replica---a train of AFC echoes---appears after 1/$\Delta$ = 5 $\mu$s storage time \AD{as described in Fig. \ref{fig:multiplexing}(a)}. \AD{Due to the limited optical depth of the crystal, not all the input light is absorbed. Thus, in AFC storage experiments, the AFC echo is generally accompanied by non-absorbed (transmitted) light.} The AFC efficiency, $\eta$, is defined as the ratio between the intensity of the AFC echo and the input pulse -- in our case around $1\%$ \cite{Das}. This value agrees with the theoretical storage efficiency $\eta_{theo}=0.98\pm0.11 \%$, estimated from the Gaussian-shaped AFC peaks using $\eta_{theo} = e^{-\tilde{d}}\tilde{d}^{2}e^{-7/F^{2}}e^{-d_{0}}$ \cite{Gisin2008}. Here, $\tilde{d} = d_1/F$, $d_1$ is the peak absorption depth, $d_{0}$ is the background absorption depth, and F is again the finesse of the AFC. \AD{An illustration of a 1 MHz-wide AFC with a finesse of 2, tailored for a 5 $\mu$s storage time, is presented in Fig. \ref{fig:multiplexing}(c), along with the relevant absorption parameters $d_{0}$ and $d_{1}$}. The reduced efficiency in our experiment is due to the imperfect optical pumping caused by technical issues such as finite laser linewidth and vibration of the cryostat, which can be especially significant for AFCs with $\mu$s long storage times, as well as small optical depth. Note that the latter can be overcome using an impedance-matched cavity \cite{impedance1, duranti2023efficient}.

\subsection{Simultaneous storage of different spectral modes}

In order to create spectrally multiplexed AFCs, we utilize two phase modulators (PMs) in series, where one of the PMs generates sidebands spaced by the desired frequency interval between neighboring AFCs while the other PM creates AFCs within each of these frequency bands. Both operations are performed using serrodyne optical phase modulation \cite{neil_tmln, saglamyurek2016multiplexed, serrodyne1}. In this way, we program and create simultaneously three 1 MHz-wide AFCs with 4, 6, and 8 $\mu$s of storage time, centered at 0, 50, and 100 MHz frequency detuning within the inhomogeneous absorption linewidth of the thulium ions. As shown in Fig. \ref{fig:multiplexing}(b), each AFC receives an optical pulse whose central frequency is matched to that of the AFC and re-emits the respective echo at the pre-programmed storage time (different colors in Fig. \ref{fig:multiplexing}(b) represent the three different frequency detunings). The spectral read-out of the individual AFC re-emissions is implemented using a filter cavity at the output of the crystal and by varying the resonant frequency of the cavity to spectrally match each frequency mode.

\section{Discussion and Conclusion}

\AD{In summary, we investigate the spectroscopic properties of a 1$\%$ thulium-doped ytterbium gallium garnet (Tm:YGG) crystal at temperatures as low as 500 mK. Our measurements reveal millisecond-long optical coherence times and Zeeman level lifetimes extending into the hundreds of seconds. We investigate and discuss plausible reasons for spectral diffusion on short and long timescales. We demonstrate a multimode optical memory both in the temporal and spectral domains with microsecond-long storage time. With an optical $T_2$ of 1.1 ms in Tm:YGG, it stands as the third-longest reported, surpassed only by Er:YSO, and Eu:YSO. However, Er:YSO suffers from the strong paramagnetism of Er$^{3+}$ ions, therefore requiring a high magnetic field of a few Tesla to show good properties \cite{4msT2Er,Bottger}. On the other hand, Eu:YSO suffers from a more complicated hyperfine structure, the presence of two natural isotopes with similar abundance, and a technologically inconvenient transition wavelength \cite{6msT2Eu}. For quantum repeaters, the importance of a long $T_2$ time is always needed, regardless of the degree of freedom used for multiplexing. This is because an extended optical $T_2$ time enables longer storage of photonic quantum states in optically excited coherence, resulting in longer elementary quantum repeater link lengths, which, in turn, reduces the number of Bell state measurements required to connect such links \cite{Multiplexing}. In addition, a long optical storage time reduces the deadtime of a quantum memory in which quantum information is stored in terms of spin coherence \cite{AFC_spinwave}. This results in an increased throughput of a repeater-based quantum communication link \cite{Das}.} 

\AD{The primary advantage of the Tm$^{3+}$-doped system for quantum memory is the operating wavelength in the near-infrared because it is easily accessible with laser diodes. Also, Tm$^{3+}$ is the only rare earth that has a nuclear spin of I = $\frac{1}{2}$ with 100\% natural abundance, giving one of the simplest energy level structures necessary for quantum memory implementation. In particular, the trivalent Tm$^{3+}$-ion has an even number of electrons, it is a non-Kramers ion with no first-order electronic magnetism while still providing a suitable hyperfine structure with a GHz-level energy-level splitting that offers long-lived, optically addressable states with lower sensitivity to the magnetically induced decoherence. Merely a few hundred Gauss of magnetic field is required to achieve a long optical coherence time, rendering it a suitable candidate for long-lived quantum memories \cite{etesse2021optical}. This stands in contrast to Kramers ions, such as erbium, where a high magnetic field around a few Tesla is necessary to achieve a long optical coherence time and to reduce spectral diffusion \cite{Bottger}. Tm:YGG exhibits reduced spectral diffusion and instantaneous spectral diffusion compared to other known Tm-doped materials such as Tm:YAG and Tm:LiNbO$_3$. A detailed investigation \cite{Thiel_2014} reveals a sensitivity to excitation-induced decoherence that is lower by more than two orders of magnitude compared to Tm:YAG and Tm:LiNbO$_3$. It shows that the Tm:YGG system features the longest optical coherence lifetimes and the lowest levels of excitation-induced decoherence observed for any known thulium-doped material.} 

Our results confirm that Tm:YGG is a promising material candidate for multimode, long-lived, and AFC-based quantum memories. However, more detailed spectroscopic measurements are required to understand the material’s full potential, such as if the coherence properties can be further improved by using a different magnetic field orientation \cite{Davidson}, by growing non-birefringent YGG crystals, and by optimizing the material composition for specific quantum memory implementations. For example, it is possible to increase the inhomogeneous broadening by co-doping selected impurities or introducing static crystal strain to further increase the spectral multiplexing capacity of the memory \cite{Thiel_strain, Bottger_codoping}. \AD{While high-quality rare-earth gallium garnet crystals can be readily grown for long-established technologically important cases, such as Gd$_3$Ga$_5$O$_{12}$ (GGG) used extensively as a substrate in the semiconductor industry and Tb$_3$Ga$_5$O$_{12}$ (TGG) used in most bulk commercial optical isolators \cite{watanabe4737066characterization, nizhankovskiy2021comparison}, much less development work has been carried out for Y$_3$Ga$_5$O$_{12}$ (YGG). Discussing these efforts extends beyond the scope of this manuscript. Polarization rotation resulting from birefringence within the crystal can negatively impact the polarized interaction of light with different Tm sub-sites \cite{Tm_polarization_PRB}. This limits the ability to selectively interact only with chosen subsets of Tm ions, each of which has a different Rabi frequency.  Furthermore, since the applied external magnetic field has a different projection on each of the sub-sites, the nuclear hyperfine structure and dynamics of each sub-site will also be significantly different.  As a result, for a general orientation, only some of the Tm sub-sites will contribute to the long-lived spectral hole burning, depending on both the orientation of the magnetic field and the optical polarization. Consequently, there are only a few specific combinations of orientations of the magnetic field and polarization relative to the crystal axes that maximize the interaction, resulting in improved optical depth and memory efficiency. Additionally, the strain-induced birefringence is also indicative of the presence of defects in the grown materials. Hence, growing improved low-strain crystals without the observed birefringence can be expected to result in further improved optical coherence time as well as nuclear spin-state lifetimes.} In parallel, optical pumping strategies must be optimized \cite{jobez2016, das_quadratic}. Furthermore, technical developments such as sub-kHz laser linewidth stabilization; isolation of the crystal against cryostat vibrations; and enhanced light-matter interaction using an impedance-matched cavity are needed for the creation of an efficient, highly multimode and long-lived optical quantum memory that enables long-distance quantum communication.

\section*{Acknowledgments}

The authors thank G. C. Amaral, N. Alfasi, and T. Chakraborty for their experimental help. We acknowledge funding through the Netherlands Organization for Scientific Research, and the European Union's Horizon 2020 Research and Innovation Program under Grant Agreement No. 820445 and project name Quantum Internet Alliance (QIA). This material is based in part on research at Montana State University sponsored by the Air Force Research Laboratory under agreement number FA8750-20-1-1004.
\\

\textbf{Current affiliations:} The current affiliations of some of the authors are the following:
Mohsen Falamarzi Askarani: Xanadu, Toronto, ON M5G 2C8, Canada; Jacob H. Davidson: National Institute of Standards \& Technology, 325 Broadway, Boulder, CO 80305, USA; Sara Marzban: MESA+ Institute for Nanotechnology, University of Twente, 7500 AE Enschede, The Netherlands; Joshua Slater: Q*Bird b.v., Delftechpark 1, 2627 XJ, Delft.

\bibliographystyle{apsrev4-1}
\bibliography{Tm-storage}
\end{document}